\documentclass[10pt, indentfirst]{article}

\usepackage{amsmath}
\usepackage{amsfonts}   
\usepackage{amssymb}    
\usepackage{graphicx}   
\usepackage{xcolor}
\usepackage{bm}
\usepackage{secdot}
\usepackage{mathptmx}
\usepackage{float}
\usepackage{titlesec}
\titleformat{\section}{\large\bfseries}{\thesection.}{1em}{}
\titleformat*{\subsection}{\normalsize\bfseries}

\usepackage[utf8]{inputenc}
\usepackage{textcomp}
\usepackage[hang,flushmargin]{footmisc} 
\usepackage[numbers,sort&compress]{natbib}
\usepackage[hidelinks]{hyperref}

\usepackage{indentfirst}
\usepackage{fancyhdr}
\usepackage{caption} 
\DeclareMathOperator\erf{erf}

\begin{document}
\setcounter{page}{1}

\title{
	\small Volume XXX (Year) \textcolor{blue}{https://doi.org/10.6028/jres.XXX.XXX} \\ 
	\large Journal of Research of National Institute of Standards and Technology \\
	\LARGE \hrulefill\\
	\Huge{\textit{Distributed Error-Function Roughness in Refl1d Reflectometry Fitting Program}}
}
\date{\vspace{-7ex}}

\vspace*{-111pt}{\let\newpage\relax\maketitle}

\maketitle

\begin{flushleft} \normalsize \textbf{Brian B. Maranville$^1$, Aaron Green$^2$, and Paul A. Kienzle$^1$} \\
	\vspace{9pt} 
	\normalsize $^1$National Institute of Standards and Technology, \\  
	Gaithersburg, MD, 20899 \\
	
	\vspace{9pt} 
	\normalsize $^2$University of Maryland, Dept. of Computer Science, \\  
	College Park, MD, 20742 \\
	
	\vspace{15pt} 
	\footnotesize \textcolor{blue}{brian.maranville@nist.gov}\\
	\footnotesize \textcolor{blue}{agreen991@gmail.com}\\
	\footnotesize \textcolor{blue}{paul.kienzle@nist.gov}\\
	\vspace{8pt} 
	\textbf{Software DOI:} https://doi.org/XXXXX/XXXXX \\ 
	\vspace{8pt} 
	\textbf{Software Version:} X.X \\
	\vspace{8pt} 
	\vspace{8pt} 
	\textbf{Accepted:} \today \\
	\vspace{8pt} 
	\textbf{Published:} \today \\ 
	\vspace{8pt} 
	\textcolor{blue}{https://doi.org/10.6028/jres.XXX.XXX}
\end{flushleft}
\vspace{-15pt}
\LARGE \hrulefill

\setlength{\abovedisplayskip}{19pt}
\setlength{\belowdisplayskip}{19pt}
\setlength{\abovedisplayshortskip}{9pt}
\setlength{\belowdisplayshortskip}{19pt}

\raggedright
\setlength\parindent{16pt}

\pagestyle{fancy}
\fancyhead{}
\fancyfoot{} 
\fancyfoot[c]{\thepage}
\fancyfoot[r]{\textcolor{blue}{https://doi.org/10.6028/jres.XXX.XXX}}
\chead{
 \small Volume XXX (Year) \textcolor{blue} 
 {https://doi.org/10.6028/jres.XXX.XXX} \\ 
 \large Journal of Research of National Institute of Standards and Technology \\} 

\section{Summary}
\label{sec:summary}

\normalsize The Refl1d \cite{kirby_phase-sensitive_2012} program is used for modeling and fitting data from neutron and X-ray reflectometry
experiments.  The model of the (thin-film) samples is typically constructed of discrete layers of different scattering-length
densities (SLD).  Interlayer roughness is represented as an error-function transition from one layer to the next.

Previous versions of the software truncated this error-function at the next interface.  This strategy has the advantage of preventing
layers with unbounded effective extent, but it can also result in SLD depth profiles that do not conform to the physical expectations of the users (such as introducing sharp transitions) whenever the layer roughness approaches the thickness of the layer.

In this article we introduce a new version of the software in which the option is provided to extend the roughness of each layer over the
entire structure; the resulting SLD profiles often more closely resemble the physical models intended by the user.  Most importantly 
no sharp transitions are introduced by truncating the roughness, when a smooth transition is often desired when adding rough layers together.

The algorithm is straightforward: for each layer of in the model, the boundaries on each side are converted to an error-function with
a width equal to the the roughness specified for that interface, offset from zero SLD by one-half the asymptotic height of the error function
and multiplied by the one-half the SLD for that layer, as specified in 
Eq. \ref{eq:erf},
\begin{equation}
\label{eq:erf}
\rho(z) = \sum_{i=0}^{N}  \left(\frac{\textrm{SLD}_{i+1} - \textrm{SLD}_{i}}{2} \right)  \left(\erf\left(\frac{z-Z_i}{\sigma_i/\sqrt{2}}\right) + 1\right)
\end{equation}
where SLD$_i$ refers the the defined SLD for layer $i$, and  $\sigma_i$ and $Z_i$ are the roughness and position (respectively) of the interface between layers $i$ and $i+1$. The sum is carried out over the $N$ interfaces in the model, including those with the semi-infinite media above, ($i=N$) and below, ($i=0$).
The rendered profile $\rho(z)$ then has contributions from all the interfaces at all positions

Note that the zero-roughness case reverts to a step form, and that 
two error functions are being added together with the same center at each interface (one for the lower boundary of a layer, and one for the upper boundary of the layer below) leading to a smooth transition between the SLD of the layers.

We have chosen to truncate the rendering at the first point greater than three-$\sigma$ from all the interfaces in the profile, in both the positive and negative directions along $z$.  This truncation
has a minimal effect on the calculated reflectivity, as at this point the error-functions have practically reached their asymptotic limits. 

\begin{figure}
\includegraphics[width=\linewidth]{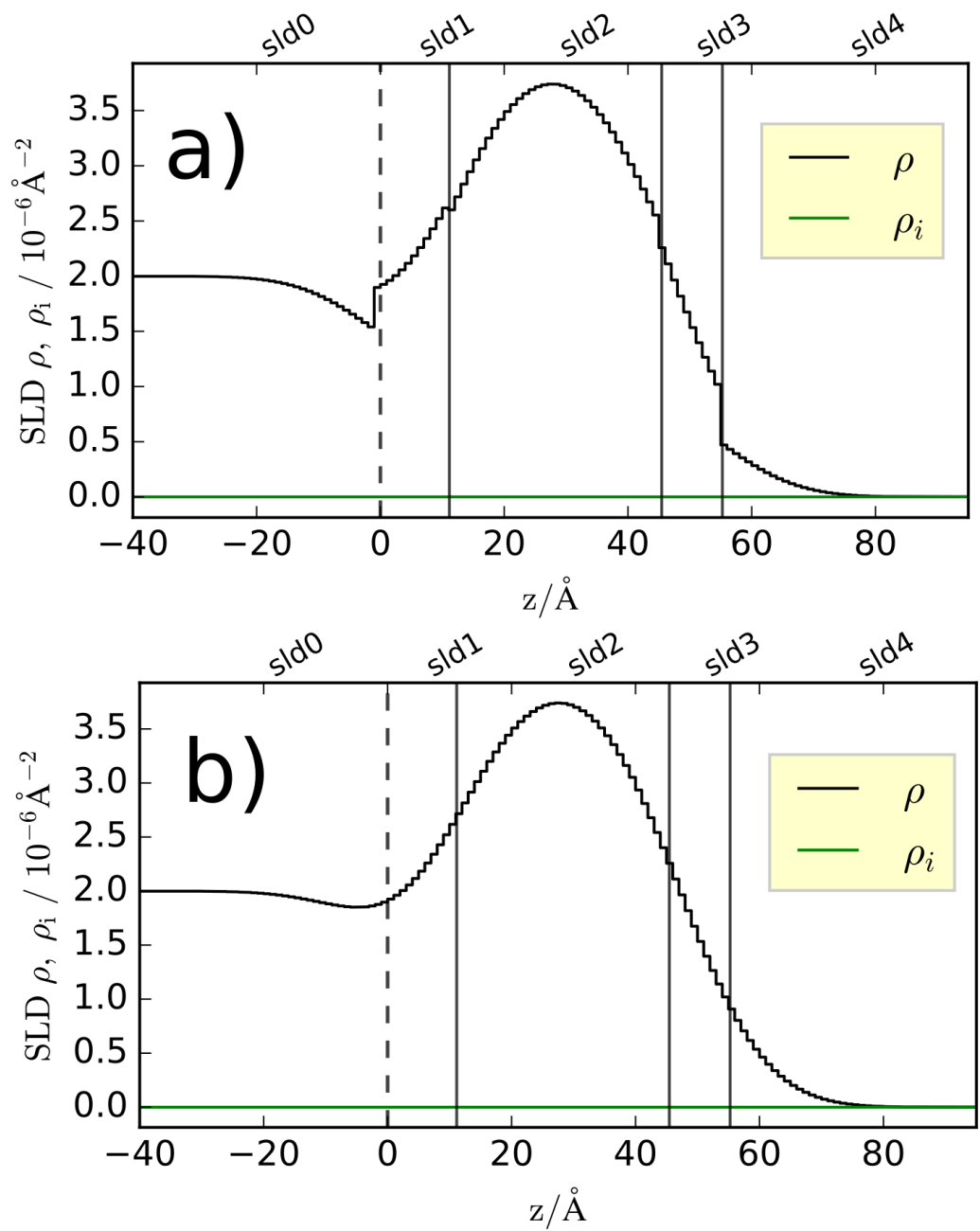}
\caption{Profile examples for a) old (truncating) algorithm for interface, and 
b) new extended interfaces.  The profile is defined in Table \ref{table:profile}}
\label{fig:profiles}
\end{figure}

An example of the change in SLD rendering is provided in Fig. \ref{fig:profiles}, for a profile defined as 
\begin{table}[H]
	\caption{Definition of example profile, where incident medium is the last layer and roughness is defined as the interface between that layer and the next, so that the roughness of the last layer is meaningless (and set to zero)}
	\label{table:profile}
	\centering
	\begin{tabular}{l | l | l | l}
		Thickness $(\AA)$ & SLD $(\AA^{-2})$ & SLD (imag.) $(\AA^{-2})$ & Roughness $(\AA)$ \\ \hline
		0 &	2 &	0 &	10 \\
		11.15&	1&	0&	10\\
		34.3&	4&	0&	10\\
		9.8&	1&	0&	10\\
		0&	0&	0&	0
	\end{tabular}
\end{table}

\section{Software Specifications}
\label{sec:softwarespec}


\begin{table}[H]
	\centering
	\small
	\def\arraystretch{1.5}
	\begin{tabular}{|l|l|}
		\hline
		\textbf{NIST Operating Unit} & NIST Center for Neutron Research \\ \hline
		\textbf{Category} & Reflectometry Modeling and Fitting  \\ \hline
		\textbf{Targeted Users} & Researchers with Reflectometry Data \\ \hline
		\textbf{Operating Systems} & Microsoft Windows, Apple OS X, Linux \\ \hline
		\textbf{Programming Language} & Python \\ \hline
		\textbf{Inputs/Outputs} & Inputs: Reflectometry data, starting model \\
		 & Outputs: Optimized model with parameter confidence intervals and correlations \\ \hline
		\textbf{Documentation} & \textcolor{blue}{http://refl1d.readthedocs.io/en/latest/} \\ \hline
		\textbf{Accessibility} & N/A \\ \hline
		\textbf{Disclaimer} & \textcolor{blue}{https://www.nist.gov/director/licensing} \\
		\hline
	\end{tabular}
\end{table}

\section{Methods for Validation}
\label{sec:methods}
Because the interface functions now extend over the entire profile, a new type of unphysical rendering can occur: if 
a layer has dissimilar roughnesses on its two boundaries, and one or both of those roughnesses are large compared to the layer 
thickness, an oscillation can be created in the rendered profile that is probably not intended by the user (arises only from the 
application of the assymmetric roughnesses).
This situation can be avoided by applying constraints that require that the roughness vary slowly compared to the roughness width
across the film, which is a reasonable physical constraint to apply in most cases. 

The changes described do not alter the calculation kernel used in the software, only the algorithm for rendering an SLD profile based
on the users' inputs.  This profile is then used at the input to the calculation kernel for all modeling, so validation is needed only for
the profile rendering.  Users typically manually inspect the rendered profile as part of the modeling process, providing a sanity check
on the new method.  \\


\vspace{20pt}

\noindent{\textbf{Acknowledgments}}\\
\noindent This research (as part of the SHIP program) was carried out with support from the Center for High Resolution Neutron Scattering, a partnership between the National Institute of Standards and Technology and the National Science Foundation under Agreement No. DMR-1508249. \\
{
	\footnotesize
\bibliographystyle{jresnist}
\bibliography{References-JRes}} 


\vspace{20pt} 

\noindent\textit{\textbf{About the authors:} Brian Maranville and Paul Kienzle are scientists at the NIST Center for Neutron Research. Aaron Green participated in this research as part of the Summer High School Internship Program (SHIP) at NIST in 2016. The National Institute of Standards and Technology is an agency of the U.S. Department of Commerce.} \\

\end{document}